\documentclass[fleqn,3p,onecolumn,11pt,nofootinbib,showkeys,showpacs]{revtex4-1}

\usepackage[left=2cm,right=2cm,top=2cm,bottom=2cm]{geometry}
\usepackage[small,justification=centerlast]{caption}
\usepackage{graphicx}
\usepackage{bm}
\usepackage{mathrsfs}
\usepackage[latin1]{inputenc}
\usepackage{stmaryrd}
\usepackage{array}
\usepackage{psfrag}
\usepackage{dsfont}
\usepackage{epsfig}
\usepackage{titletoc}
\usepackage{float}   %  for figures
\usepackage{wrapfig} %  for figures
\usepackage{amsmath}
\usepackage{amssymb,stmaryrd}
\usepackage{psfrag}
\usepackage{cancel}
\usepackage[dvips]{feynmp}
\usepackage{upgreek}
\usepackage{subfig}
\usepackage{tabularx}
\usepackage{xcolor}
\usepackage{units}

\usepackage{textfit} %\scaletoheight{0.05cm}{very small text}
\usepackage{hyperref}

\renewcommand{\eqref}[1]{\mbox{Eq.~(\ref{#1})}}

\newcommand{\secref}[1]{\mbox{Sec.~\ref{#1}}}

\begin{document}

\title{Classical kinematics for isotropic, minimal \\ Lorentz-violating fermion operators}

\author{M. Schreck}\email{mschreck@indiana.edu}
\affiliation{Indiana University Center for Spacetime Symmetries, \\ Indiana University, Bloomington, Indiana 47405-7105}

\begin{abstract}
In this article a particular classical, relativistic Lagrangian based on the isotropic fermion sector of the Lorentz-violating (minimal) Standard-Model Extension
is considered. The motion of the associated classical particle in an external electromagnetic field is studied and the evolution of its spin, which is
introduced by hand, is investigated. It is shown that the particle travels along trajectories that are scaled versions of the standard ones.
Furthermore there is no spin precession due to Lorentz violation, but the rate is modified at which the longitudinal and transverse spin components
transform into each other. This demonstrates that it is practical to consider classical physics within such an isotropic Lorentz-violating framework
and it opens the pathway to study a curved background in that context.
\end{abstract}

\maketitle

\newpage%%tmp
\section{Introduction}

Since violations of {\em CPT} symmetry and Lorentz invariance were shown to appear in the context of string theory \cite{Kostelecky:1988zi,Kostelecky:1991ak,
Kostelecky:1994rn,oai:arXiv.org:hep-th/9605088}, the interest in exploring a possible violation of this fundamental symmetry in nature has
grown steadily. Subsequently such a violation was also found to occur in loop quantum gravity \cite{Gambini:1998it,Bojowald:2004bb}, models of
noncommutative spacetimes \cite{Carroll:2001ws}, spacetime foams models \cite{Klinkhamer:2003ec,Bernadotte:2006ya}, and in spacetimes endowed
with a nontrivial topology \cite{Klinkhamer:1998fa,Klinkhamer:1999zh}.
Therefore it can be considered as a window to physics at the Planck scale. A further boom creating a new field of research took place when the
minimal Standard-Model Extension (SME) was established \cite{Colladay:1998fq}. The latter provides a powerful effective framework for describing
Lorentz violation for energies much smaller than the Planck scale.

Since then the field has been developing largely in both experimental searches for Lorentz violation and the study of theoretical aspects. There has been a broad
experimental search for Lorentz violation (see the data tables \cite{Kostelecky:2008ts} and references therein) and there are ongoing studies on the
properties of quantum field theories
based on the SME \cite{Kostelecky:2000mm,oai:arXiv.org:hep-ph/0101087,Casana-etal2009,Casana-etal2010,Klinkhamer:2010zs,Schreck:2011ai,
Cambiaso:2012vb,Colladay:2014dua,Maniatis:2014xja,Schreck:2013gma,Schreck:2013kja,Cambiaso:2014eba,Schreck:2014qka}. Recently,
the nonminimal versions of the SME including all higher-dimensional operators of the photon, fermion, and neutrino sector have been constructed
as well \cite{Kostelecky:2009zp,Kostelecky:2011gq,Kostelecky:2013rta}.

Although the SME seems to work very well in flat spacetime, certain issues arise when it is coupled to gravitational fields. Around ten years ago
a no-go theorem was proven stating that an explicitly Lorentz-violating field theory cannot be coupled to gravity consistently, because this leads to
incompatibilities with the Bianchi identities \cite{Kostelecky:2003fs}.\footnote{Besides, note that certain tensions with the generalized second law
of black-hole thermodynamics may occur when particular Lorentz-violating theories are coupled to a black-hole gravitational background. The reason
is the multiple-horizon structure, e.g., for photons that arises in such frameworks \cite{Dubovsky:2006vk,Eling:2007qd,Betschart:2008yi,Kant:2009pm}.} A
coupling is only possible if Lorentz invariance is violated spontaneously, e.g., in a Bumblebee model \cite{Kostelecky:1988zi,Kostelecky:1989jp,
Kostelecky:1989jw,Kostelecky:2000mm,Kostelecky:2003fs,Bailey:2006fd,Bluhm:2008yt}.

Note that the incompatibilities mentioned were found in the context of Riemann-Cartan spacetimes, i.e., spacetimes endowed with the Riemannian
concept of curvature including torsion. An alternative approach to considering Lorentz violation in gravitational backgrounds is to change the
fundamental geometrical concept. Hence instead of Riemann-Cartan geometry one might be tempted to use Finsler geometry
\cite{Finsler:1918,Cartan:1933,Matsumoto:1986,Antonelli:1993,Bao:2000,Kozma:2003,Bao:2004,Bucataru:2007} as the basis of a theory of gravity.
Geometrical quantities in Finsler spaces such as curvature do not only depend on the particular point considered in the space but also on the angle
that a given line element encloses with an inherent direction in this space. Finsler spaces rest on more general length functionals,
whereby they can be considered as Riemannian spaces without the quadratic restriction \cite{Chern:1996}.

For this reason Finsler geometry may be a natural framework to describe preferred directions in a curved spacetime, i.e., Lorentz violation in the
presence of gravity. Lately plenty of work has been done to identify Finsler spaces linked to certain cases of the SME fermion
sector, which includes studies of the minimal \cite{Kostelecky:2010hs,Kostelecky:2011qz,Kostelecky:2012ac,Colladay:2012rv,Russell:2015gwa} and also the
nonminimal sector \cite{Schreck:2014hga}. In the current article isotropic subsets of the minimal fermion sector will be investigated. We will obtain
the corresponding Finsler structure and address certain physical problems such as the propagation of a classical, relativistic, pointlike particle in
the Lorentz-violating background and the time evolution of particle spin.

The paper is organized as follows. In \secref{sec:isotropic-dispersion-laws-minimal} all isotropic coefficients of the minimal SME fermion
sector are identified and the corresponding dispersion relations are computed. In \secref{sec:construction-finsler-structure} a generic
isotropic dispersion relation is considered and its associated classical, relativistic Lagrangian is derived, which is then promoted to a
Finsler structure. Section \ref{sec:particle-electromagnetic-field} is dedicated to studying the physics of the classical Lagrangian obtained.
First of all the motion of the classical particle in an electromagnetic field will be analyzed. Besides, the interest also lies in the behavior of
particle spin, which is introduced by hand and treated with the Bargmann-Michel-Telegdi (BMT) equation \cite{Bargmann:1959gz}. Finally the results
are summarized and discussed in \secref{sec:discussion}. Throughout the paper natural units with $c=\hbar=1$ are used unless otherwise stated.

%---------------------------------------------------------------------------------------------------
\section{Isotropic dispersion laws in the minimal fermion sector}
\label{sec:isotropic-dispersion-laws-minimal}
%---------------------------------------------------------------------------------------------------

The intention of the current section is to find all isotropic dispersion relations of the minimal SME fermion sector. The full action including both
minimal and nonminimal contributions reads as \cite{Kostelecky:2013rta}
\begin{subequations}
\begin{align}
S&=\int_{\mathbb{R}^4} \mathrm{d}^4x\,\mathcal{L}\,,\quad \mathcal{L}=\frac{1}{2}\overline{\psi}\left(\gamma^{\mu}\mathrm{i}\partial_{\mu}-m_{\psi}\mathds{1}_4+\widehat{\mathcal{Q}}\right)\psi+\text{H.c.}\,, \\[2ex]
\widehat{\mathcal{Q}}&=\mathrm{i}\left(\widehat{c}^{\,\mu\alpha_1}\gamma_{\mu}+\widehat{d}^{\mu\alpha_1}\gamma_5\gamma_{\mu}+\widehat{e}^{\,\alpha_1}\mathds{1}_4+\mathrm{i}\widehat{f}^{\,\alpha_1}\gamma_5+\frac{1}{2}\widehat{g}^{\,\mu\nu\alpha_1}\sigma_{\mu\nu}\right)\partial_{\alpha_1} \notag \\
&\phantom{{}={}}-\left(\widehat{m}\mathds{1}_4+\mathrm{i}\widehat{m}_5\gamma_5+\widehat{a}^{\mu}\gamma_{\mu}+\widehat{b}^{\mu}\gamma_5\gamma_{\mu}+\frac{1}{2}\widehat{H}^{\mu\nu}\sigma_{\mu\nu}\right)\,.
\end{align}
\end{subequations}
Here $\psi$ is a Dirac spinor field, $\overline{\psi}\equiv \psi^{\dagger}\gamma^0$ its Dirac conjugate, and $m_{\psi}$ is the fermion mass.
The $\gamma^{\mu}$ for $\mu=0\dots 3$ are the standard Dirac matrices obeying the Clifford algebra $\{\gamma^{\mu},\gamma^{\nu}\}=2\eta^{\mu\nu}\mathds{1}_4$
and $\mathds{1}_4$ is the unit matrix in spinor space.
The operator $\widehat{\mathcal{Q}}$ is a collection of all minimal and nonminimal Lorentz-violating composite operators in the pure fermion
sector. All fields and operators are defined in Minkowski spacetime with the metric $(\eta_{\mu\nu})=\mathrm{diag}(1,-1,-1,-1)$.

The terminology to denoting Lorentz-violating operators is chosen according to Table 1 of \cite{Kostelecky:2013rta}. Each operator is characterized
by a couple of free Lorentz indices whose number ranges from 0 to 2. These indices control the spin behavior of the corresponding operator. Upon
decomposition of $\widehat{\mathcal{Q}}$ into the Dirac bilinears according to Eq.~(2) in the latter reference the operators are grouped into scalars,
vectors, and second-rank tensors.
Additionally, in momentum space these are split into momenta and Lorentz-violating component coefficients, cf. Eqs.~(5), (6) in
\cite{Kostelecky:2013rta}. Their transformation properties with respect to (proper and improper) observer Lorentz transformations
and charge conjugation are stated in Table 1 of \cite{Kostelecky:2013rta} as well. Both the scalar $\widehat{m}$ and the pseudoscalar operator $\widehat{m}_5$
only appear in the nonminimal sector, i.e., the analysis will be restricted to the vector operators $\widehat{a}^{\mu}$, $\widehat{b}^{\mu}$,
$\widehat{c}^{\,\mu}$, $\widehat{d}^{\mu}$, the scalar operators $\widehat{e}$, $\widehat{f}$, and the tensor operators $\widehat{g}^{\mu\nu}$,
$\widehat{H}^{\mu\nu}$.
The following calculations will be based on Eq.~(39) of \cite{Kostelecky:2013rta}, which gives the general dispersion relation of the SME fermion
sector including all minimal and nonminimal contributions. The dispersion relation
involves the operators $\widehat{\mathcal{S}}$, $\widehat{\mathcal{P}}$, $\widehat{\mathcal{V}}$, $\widehat{\mathcal{A}}$, $\widehat{\mathcal{T}}^{\mu\nu}$
defined by Eqs.~(2), (7) and $\widehat{\mathcal{S}}_{\pm}$, $\widehat{\mathcal{V}}^{\mu}_{\pm}$, $\widehat{\mathcal{T}}^{\mu\nu}_{\pm}$
given by Eq.~(35) in the latter reference.

First of all the vector operators $\widehat{a}^{\mu}\equiv a^{(3)\mu}$ and $\widehat{b}^{\mu}\equiv b^{(3)\mu}$ shall be considered. They are contained in
$\widehat{\mathcal{V}}^{\mu}$ and $\widehat{\mathcal{A}}^{\mu}$, respectively, and they contribute to $\widehat{\mathcal{V}}^{\mu}_{\pm}$.
For $\widehat{\mathcal{S}}_{\pm}=-m_{\psi}$, $\widehat{\mathcal{V}}^{\mu}_{\pm}=p^{\mu}+\widehat{\mathcal{V}}^{\mu}$, and $\widehat{\mathcal{T}}_{\pm}^{\mu\nu}=0$
the modified fermion dispersion relation results in
\begin{equation}
\label{eq:dispersion-law-vector-coefficients-1}
p^2+2p\cdot \widehat{\mathcal{V}}+\widehat{\mathcal{V}}^2-m_{\psi}^2=0\,,
\end{equation}
with the fermion four-momentum $p^{\mu}$.
Setting $\widehat{\mathcal{V}}^{\mu}=-\widehat{a}^{\mu}$ the second term on the left-hand side of the latter equation cannot be isotropic for
any choice of $\widehat{a}^{\mu}$ besides $(a^{(3)\mu})=(a^{(3)0},0,0,0)^T$. The corresponding dispersion relation is then given by
\begin{equation}
(p_0)^+=a^{(3)0}+\sqrt{\mathbf{p}^2+m_{\psi}^2}\,,
\end{equation}
where $\mathbf{p}$ is the particle three-momentum. Here $(p_0)^+$ denotes the positive-energy dispersion law.
This result is encoded in Eq.~(94) of \cite{Kostelecky:2013rta}. Note that a nonzero coefficient $a^{(3)0}$ just leads to an unobservable shift
of the particle energy, which reminds us of the fact that the coefficients $a^{(3)\mu}$ can be removed by a phase redefinition
\cite{Kostelecky:2013rta}.
As a next step the operator $\widehat{b}^{\mu}$ is considered. From $\widehat{S}_{\pm}=-m_{\psi}$, $\widehat{\mathcal{V}}^{\mu}_{\pm}=p^{\mu}\pm\widehat{\mathcal{A}}^{\mu}$,
and $\widehat{\mathcal{T}}^{\mu\nu}_{\pm}=0$ we obtain:
\begin{equation}
\label{eq:dispersion-law-vector-coefficients-2}
(p^2+2p\cdot\widehat{\mathcal{A}}+\widehat{\mathcal{A}}^{\,2})(p^2-2p\cdot\widehat{\mathcal{A}}+\widehat{\mathcal{A}}^{\,2})-2m_{\psi}^2(p^2-\widehat{\mathcal{A}}^{\,2})+m_{\psi}^4=0\,.
\end{equation}
For $\widehat{A}^{\mu}=-\widehat{b}^{\mu}$ the term $p\cdot \widehat{\mathcal{A}}$ can only be isotropic, if $(b^{(3)\mu})=(b^{(3)0},0,0,0)^T$.
Then there are two different dispersion relations that read as
\begin{equation}
\label{eq:dispersion-relation-on-shell-b0}
(p_0)^+_{1,2}=\sqrt{\mathbf{p}^2+m_{\psi}^2+(b^{(3)0})^2\pm 2|b^{(3)0}||\mathbf{p}|}\approx \sqrt{\mathbf{p}^2+m_{\psi}^2}\left(1\pm |b^{(3)0}|\frac{|\mathbf{p}|}{\mathbf{p}^2+m_{\psi}^2}\right)\,.
\end{equation}
The expansion here and all subsequent ones are understood to be valid for a sufficiently small Lorentz-violating coefficient.
Due to Lorentz violation the energies of fermion states with different spin projections are no longer degenerate. This behavior resembles a
birefringent vacuum for the photon sector. 

The situation is slightly similar for the vector operators $\widehat{c}^{\,\mu}\equiv c^{(4)\mu\alpha_1}p_{\alpha_1}$ and $\widehat{d}^{\mu}\equiv d^{(4)\mu\alpha_1}p_{\alpha_1}$
being comprised of second-rank tensor coefficients that are contracted with one additional four-momentum. We consider
$\widehat{\mathcal{V}}^{\mu}=c^{(4)\mu\alpha_1}p_{\alpha_1}$ at first. To end up with an isotropic dispersion relation, the coefficients $c^{(4)\mu\alpha_1}$
must be chosen such that $p\cdot\widehat{\mathcal{V}}=p_{\mu}c^{(4)\mu\alpha_1}p_{\alpha_1}$ in \eqref{eq:dispersion-law-vector-coefficients-1} is
isotropic. This is only the case if all off-diagonal components vanish and $c^{(4)11}=c^{(4)22}=c^{(4)33}$. Since $c^{(4)\mu\alpha_1}$ is traceless,
that heavily restricts the possibilities of choices for the coefficients, with only
one remaining:
\begin{equation}
\label{eq:isotropic-choice-cmunu}
(c^{(4)\mu\nu})=c^{(4)00}\,\mathrm{diag}\left(1,\frac{1}{3},\frac{1}{3},\frac{1}{3}\right)\,.
\end{equation}
Then even $p_{\mu}c^{(4)\mu\alpha_1}p_{\alpha_1}=0$, which makes the dispersion relation manifestly isotropic. The coefficients $d^{(4)\mu\alpha_1}$
behave in a similar manner.
Setting $\widehat{\mathcal{A}}^{\mu}=d^{(4)\mu\alpha_1}p_{\alpha_1}$, the expression $p\cdot\widehat{\mathcal{A}}=p_{\mu}d^{(4)\mu\alpha_1}p_{\alpha_1}$
in \eqref{eq:dispersion-law-vector-coefficients-2} must be isotropic. With an analogous argument this leads to
\begin{equation}
\label{eq:isotropic-choice-dmunu}
(d^{(4)\mu\nu})=d^{(4)00}\,\mathrm{diag}\left(1,\frac{1}{3},\frac{1}{3},\frac{1}{3}\right)\,.
\end{equation}
For the choice of \eqref{eq:isotropic-choice-dmunu} it can be checked that $p_{\mu}d^{(4)\mu\alpha_1}p_{\alpha_1}=0$
resulting in an isotropic dispersion relation. Now the modified dispersion laws in case of nonvanishing coefficients $c^{(4)00}$ and
$d^{(4)00}$ read as follows:
\begin{align}
\label{eq:dispersion-relation-on-shell-c00-d00}
(p_0)_{1,2}^+&=\frac{\sqrt{\left[3+(2-c^{(4)00})c^{(4)00}+(d^{(4)00})^2\right]^2\mathbf{p}^2+9\left[(1+c^{(4)00})^2-(d^{(4)00})^2\right]m_{\psi}^2}\pm 4d^{(4)00}|\mathbf{p}|}{3\left[(1+c^{(4)00})^2-(d^{(4)00})^2\right]} \notag \\
&\approx \sqrt{\mathbf{p}^2+m_{\psi}^2}\left(1-c^{(4)00}\frac{(4/3)\mathbf{p}^2+m_{\psi}^2}{\mathbf{p}^2+m_{\psi}^2}\right)\pm \frac{4}{3}d^{(4)00}|\mathbf{p}|\,.
\end{align}
For $d^{(4)00}=0$ there is a single dispersion relation for both spin projections of the fermion. At first order in $c^{(4)00}$
(and for $m_{\psi}=0$) this modification corresponds to the modification appearing in the isotropic, {\em CPT}-even extension of the photon sector.
That is reasonable, since both sectors are related by a coordinate transformation (see \cite{Altschul:2006zz} and references therein). The result is
confirmed by Eq.~(95) in \cite{Kostelecky:2013rta}. For $d^{(4)00}\neq 0$ there exist two distinct isotropic dispersion relations.

The next step is to consider the scalar operators $\widehat{e}\equiv e^{(4)\alpha_1}p_{\alpha_1}$ and $\widehat{f}\equiv f^{(4)\alpha_1}p_{\alpha_1}$.
For the operator $\widehat{e}$ it holds that $\widehat{S}=\widehat{e}$, $\widehat{S}_{\pm}=-m_{\psi}+\widehat{e}$,
$\widehat{\mathcal{V}}^{\mu}_{\pm}=p^{\mu}$, and $\widehat{\mathcal{T}}_{\pm}^{\mu\nu}=0$, which is subsequently inserted in
Eq.~(39) of~\cite{Kostelecky:2013rta} to give
\begin{equation}
p^2-(m_{\psi}-\widehat{e}\,)^2=0\,.
\end{equation}
The latter can only be isotropic for $(e^{(4)\alpha_1})=(e^{(4)0},0,0,0)^T$ leading to the dispersion relation
\begin{equation}
(p_0)^+=\frac{\sqrt{[1-(e^{(4)0})^2]\mathbf{p}^2+m_{\psi}^2}-e^{(4)0}m_{\psi}}{1-(e^{(4)0})^2}\approx\sqrt{\mathbf{p}^2+m_{\psi}^2}-e^{(4)0}m_{\psi}\,.
\end{equation}
The result corresponds to the observation that $a_{\mathrm{eff}}^{(5)000}$ is isotropic (see Eq.~(97) in \cite{Kostelecky:2013rta})
where this effective dimension-5 coefficient also contains $e^{(4)0}$ according to the first of Eqs.~(27) in \cite{Kostelecky:2013rta}.
A similar investigation can be carried out for $\widehat{f}$ where $\widehat{\mathcal{S}}_{\pm}=-m_{\psi}\pm\mathrm{i}\widehat{\mathcal{P}}$,
which gives
\begin{equation}
p^2-(m_{\psi}^2+\widehat{f}^{\,2})=0\,.
\end{equation}
The latter result can only be isotropic for $(f^{(4)\alpha_1})=(f^{(4)0},0,0,0)^T$, whereby one obtains
\begin{equation}
(p_0)^+=\sqrt{\frac{\mathbf{p}^2+m_{\psi}^2}{1-(f^{(4)0})^2}}\approx\sqrt{\mathbf{p}^2+m_{\psi}^2}\left(1+\frac{1}{2}(f^{(4)0})^2\right)\,.
\end{equation}
Note that by a spinor redefinition the coefficients $f^{(4)\alpha_1}$ can be transferred to the $\widehat{c}^{\,\mu}$ operator
\cite{Altschul:2006ts}. In addition, the dispersion relation is only affected at second order for this particular type of coefficients.

Last but not least the tensor coefficients $\widehat{H}^{\mu\nu}\equiv H^{(3)\mu\nu}$ and $\widehat{g}^{\mu\nu}\equiv g^{(4)\mu\nu\alpha_1}p_{\alpha_1}$ will be investigated.
They are both contained in the tensor operator $\widehat{\mathcal{T}}^{\mu\nu}=\widehat{g}^{\mu\nu}-\widehat{H}^{\mu\nu}$. The special case
derived from the general dispersion relation of Eq.~(39) in \cite{Kostelecky:2013rta} by setting $\widehat{\mathcal{S}}_{\pm}=-m_{\psi}$, $\widehat{V}^{\mu}_{\pm}=p^{\mu}$
is given by:
\begin{subequations}
\begin{align}
\label{eq:dispersion-law-off-shell-tmunu}
0&=\left(m_{\psi}^2-\widehat{\mathcal{T}}_-^{\mu\nu}\widehat{\mathcal{T}}_{-,\mu\nu}\right)\left(m_{\psi}^2-\widehat{\mathcal{T}}_+^{\varrho\sigma}\widehat{\mathcal{T}}_{+,\varrho\sigma}\right)+p^4 \notag \\
&\phantom{{}={}}-2p_{\mu}\left(-m_{\psi}\eta^{\mu\nu}+2\mathrm{i}\widehat{\mathcal{T}}_-^{\mu\nu}\right)\left(-m_{\psi}\eta_{\nu\varrho}-2\mathrm{i}\widehat{\mathcal{T}}_{+,\nu\varrho}\right)p^{\varrho}\,,
\end{align}
with the convenient definition
\begin{equation}
\label{eq:definition-plus-minus-tmunu}
\widehat{\mathcal{T}}^{\mu\nu}_{\pm}\equiv\frac{1}{2}\Big(\widehat{\mathcal{T}}^{\mu\nu}\pm \mathrm{i}\widetilde{\widehat{\mathcal{T}}}^{\substack{\phantom{\mu\nu} \\ \,\mu\nu}}\Big)\,,\quad \widetilde{\widehat{\mathcal{T}}}^{\substack{\phantom{\mu\nu} \\ \,\mu\nu}}=\frac{1}{2}\varepsilon^{\mu\nu\varrho\sigma}\widehat{\mathcal{T}}_{\varrho\sigma}\,.
\end{equation}
\end{subequations}
The latter involves the dual of $\widehat{\mathcal{T}}^{\mu\nu}$, which is denoted with an additional tilde and formed by contraction of
$\widehat{\mathcal{T}}^{\mu\nu}$ with the four-dimensional Levi-Civita symbol $\varepsilon^{\mu\nu\varrho\sigma}$ where $\varepsilon^{0123}=1$.
Now \eqref{eq:dispersion-law-off-shell-tmunu} can be further simplified by using the properties of $\widehat{\mathcal{T}}^{\mu\nu}$. When adding
the operators defined in \eqref{eq:definition-plus-minus-tmunu} the dual is eliminated. Furthermore the square of $\widehat{\mathcal{T}}^{\mu\nu}$
corresponds to the square of its dual with an additional minus sign. Eventually, $\widehat{\mathcal{T}}^{\mu\nu}$ contracted with two
four-momenta vanishes:%%
\begin{equation}
\widehat{\mathcal{T}}_+^{\mu\nu}-\widehat{\mathcal{T}}_-^{\mu\nu}=\mathrm{i}\widetilde{\widehat{\mathcal{T}}}^{\substack{\phantom{\mu\nu} \\ \,\mu\nu}}\,,\quad \widehat{\mathcal{T}}^{\mu\nu}\widehat{\mathcal{T}}_{\mu\nu}=-\widetilde{\widehat{\mathcal{T}}}^{\substack{\phantom{\mu\nu} \\ \,\mu\nu}}\widetilde{\widehat{\mathcal{T}}}_{\mu\nu}\,,\quad p_{\mu}\widehat{\mathcal{T}}^{\mu\nu}p_{\nu}=0\,.
\end{equation}
The second and third relationship follow from the antisymmetry of $\widehat{\mathcal{T}}^{\mu\nu}$. By using these relations,
\eqref{eq:dispersion-law-off-shell-tmunu} can be further simplified:
\begin{equation}
\label{eq:dispersion-relation-tmunu}
p^4-2m_{\psi}^2p^2+\left(m_{\psi}^2-\frac{1}{2}\widehat{\mathcal{T}}^{\mu\nu}\widehat{\mathcal{T}}_{\mu\nu}\right)^2+\frac{1}{4}(\widehat{\mathcal{T}}^{\mu\nu}\widetilde{\widehat{\mathcal{T}}}_{\mu\nu})^2-8p_{\mu}\widehat{\mathcal{T}}_-^{\mu\nu}\widehat{\mathcal{T}}_{+,\nu\varrho}p^{\varrho}=0\,.
\end{equation}
For the tensor operator $\widehat{H}^{\mu\nu}$ the only term that may lead to anisotropy is the last one on the left-hand side of the latter equation.
By explicitly inserting $\widehat{H}^{\mu\nu}$, it can be demonstrated that no choice of the coefficients of $H^{(3)\mu\nu}$ leads to an isotropic
expression. In \cite{Kostelecky:2013rta} it was shown that only the dimension-5 coefficients $\widetilde{H}_{\mathrm{eff}}^{(5)0j0j}$
produce an isotropic dispersion law (see Eq.~(97) in
\cite{Kostelecky:2013rta}). According to the fourth of Eqs.~(27) in \cite{Kostelecky:2013rta} these effective coefficients contain $\widetilde{H}^{(5)0j0j}$ where
$\widetilde{H}^{(5)\mu\nu\alpha_1\alpha_2}$ are the dual coefficients of $H^{(5)\mu\nu\alpha_1\alpha_2}$. Furthermore, by symmetry arguments they also
comprise $d^{(4)00}$. This explains the isotropic dispersion laws of \eqref{eq:dispersion-relation-on-shell-c00-d00} following from a nonzero coefficient
$d^{(4)00}$.

Hence an isotropic dispersion relation does not exist for any of the dimension-3 component coefficients $H^{(3)\mu\nu}$. For the tensor operator $\widehat{g}^{\mu\nu}$
the situation is different. With $\widehat{\mathcal{T}}^{\mu\nu}=g^{(4)\mu\nu\alpha_1}p_{\alpha_1}$ it can be checked that there is an isotropic
dispersion relation for two different choices of coefficients. The first choice is
\begin{equation}
\label{eq:first-set-isotropic-g-coefficients}
g^{(4)123}=g^{(4)231}=g^{(4)312}\equiv g_1\,,\quad g^{(4)132}=g^{(4)213}=g^{(4)321}=-g_1\,,
\end{equation}
and all others set to zero, which results in two modified dispersion relations:
\begin{equation}
\label{eq:dispersion-relation-on-shell-gmunulambda-1}
(p_0)_{1,2}^+=\sqrt{(1+g_1^{\;\!2})\mathbf{p}^2\pm 2|g_1|m_{\psi}|\mathbf{p}|+m_{\psi}^2}\approx\sqrt{\mathbf{p}^2+m_{\psi}^2}\left(1\pm |g_1|\frac{m_{\psi}|\mathbf{p}|}{\mathbf{p}^2+m_{\psi}^2}\right)\,.
\end{equation}
The nonzero coefficients of \eqref{eq:first-set-isotropic-g-coefficients} are contained in $\widetilde{g}^{(4)0jj}_{\mathrm{eff}}$ of Eq.~(95) in \cite{Kostelecky:2013rta}
where $\widetilde{g}^{(4)\mu\nu\alpha_1}$ denotes the dual of $g^{(4)\mu\nu\alpha_1}$. According to the third of Eqs.~(27) in \cite{Kostelecky:2013rta} these effective
coefficients also contain $b^{(4)0}$, which explains the isotropic dispersion relation of \eqref{eq:dispersion-relation-on-shell-b0}. For this particular choice of
$g^{(4)\mu\nu\alpha_1}$ the last term on the left-hand side of \eqref{eq:dispersion-relation-tmunu} is isotropic.  The second choice of coefficients, which
fulfills that condition, is
\begin{equation}
\label{eq:second-set-isotropic-g-coefficients}
g^{(4)101}=g^{(4)202}=g^{(4)303}\equiv g_2\,,\quad g^{(4)011}=g^{(4)022}=g^{(4)033}=-g_2\,,
\end{equation}
and all remaining ones set to zero. This case gives rise to a single modified dispersion relation:
\begin{equation}
\label{eq:dispersion-relation-on-shell-gmunulambda-2}
(p_0)^+=\sqrt{(1+g_2^2)\mathbf{p}^2+m_{\psi}^2}\approx\sqrt{\mathbf{p}^2+m_{\psi}^2}\left(1+\frac{g_2^2}{2}\frac{\mathbf{p}^2}{\mathbf{p}^2+m_{\psi}^2}\right)\,.
\end{equation}
Note that \eqref{eq:dispersion-relation-on-shell-gmunulambda-1} comprises a modification at first order in the Lorentz-violating coefficients, whereas
the modification in \eqref{eq:dispersion-relation-on-shell-gmunulambda-2} is of second order in Lorentz violation. The term
$\widehat{\mathcal{T}}^{\mu\nu}\widehat{\mathcal{T}}_{\mu\nu}$ in \eqref{eq:dispersion-relation-tmunu} differs for both sets of component coefficients
leading to distinct dispersion relations.

To summarize, in the minimal fermion sector of the SME an isotropic dispersion relation exists for a particular choice of $a^{(3)\mu}$, $b^{(3)\mu}$,
$c^{(4)\mu\alpha_1}$, $d^{(4)\mu\alpha_1}$, $e^{(3)\alpha_1}$, $f^{(3)\alpha_1}$, and $g^{(4)\mu\nu\alpha_1}$ component coefficients. Some of
these dispersion relations depend on the spin projection of the fermion, which is the analogy of a birefringent vacuum in the photon sector.

%---------------------------------------------------------------------------------------------------
\section{Construction of the classical Lagrangian and Finsler structure}
\label{sec:construction-finsler-structure}
%---------------------------------------------------------------------------------------------------

As of now we intend to consider an isotropic modified dispersion relation of the generic form
\begin{equation}
\label{eq:modified-generic-dispersion-law}
p_0^2-\Upsilon^2\mathbf{p}^2-m_{\psi}^2=0\,,\quad (p_0)_{1,2}=\pm \sqrt{\Upsilon^2\mathbf{p}^2+m_{\psi}^2}\,,
\end{equation}
with a dimensionless parameter $\Upsilon$ where in the standard case $\Upsilon=1$.
Such a dispersion relation emerges from a particular choice of the $g^{(4)\mu\nu\alpha_1}$ coefficients, cf. \eqref{eq:dispersion-relation-on-shell-gmunulambda-2},
or for a nonvanishing $c^{(4)00}$ (see \eqref{eq:dispersion-relation-on-shell-c00-d00} by setting $d^{(4)00}=0$) when absorbing the global modification before
the square root into the fermion mass. This dispersion relation is based on the fermion Lagrangian of the SME, i.e., it is a field theory result.

In what follows, for the particular isotropic dispersion relation of \eqref{eq:modified-generic-dispersion-law} the Lagrangian $L$ shall be derived, which describes a
classical, relativistic, pointlike particle whose conjugate momentum satisfies the dispersion relation mentioned. It was shown in \cite{Kostelecky:2010hs} that such a
Lagrangian can, in principle, be obtained from five equations involving the four-momentum components $p_{\mu}$ and the four-velocity components $u^{\mu}$ of the
classical particle. One of these equations is the modified dispersion relation. Furthermore, due to parameterization invariance of the classical action along a
path the Lagrangian must be positively homogeneous of first degree in the velocity. Then it has to be of the following shape, which forms the second equation:
\begin{equation}
\label{eq:homogeneity-lagrangian}
L=-u^{\mu}p_{\mu}\,,\quad p_{\mu}=-\frac{\partial L}{\partial u^{\mu}}\,.
\end{equation}
Here $p_{\mu}$ is the conjugate momentum of the particle. Note the minus sign in the definition of the latter. If we construct a quantum-mechanical
wave packet from the quantum-theoretic free-field equations, its group velocity shall correspond to the velocity of the classical pointlike
particle:
\begin{equation}
\label{eq:group-velocities}
\frac{\partial p_0}{\partial |\mathbf{p}|}=\Upsilon^2\frac{|\mathbf{p}|}{p_0}\overset{!}{=}-\frac{|\mathbf{u}|}{u^0}\,.
\end{equation}
Because of the assumed isotropy of the Lagrangian the original three conditions, which hold for the spatial momentum components, result in only one
equation here for the magnitude $|\mathbf{p}|$ of the spatial momentum and the magnitude $|\mathbf{u}|$ of the three-velocity. This single equation
can be solved with respect to $|\mathbf{p}|$:
\begin{equation}
\frac{\Upsilon^4\mathbf{p}^2}{\Upsilon^2\mathbf{p}^2+m_{\psi}^2}=\frac{\mathbf{u}^2}{(u^0)^2} \Rightarrow |\mathbf{p}|=\frac{m_{\psi}|\mathbf{u}|}{\Upsilon\sqrt{\Upsilon^2(u^0)^2-\mathbf{u}^2}}\,.
\end{equation}
Then the zeroth four-momentum component can be expressed via the velocity as well:
\begin{equation}
\label{eq:zeroth-momentum-component}
p_0=\pm \sqrt{\Upsilon^2\mathbf{p}^2+m_{\psi}^2}=\pm \frac{\Upsilon m_{\psi}|u^0|}{\sqrt{\Upsilon^2(u^0)^2-\mathbf{u}^2}}\,.
\end{equation}
According to \eqref{eq:group-velocities} for $u^0\geq 0$ the sign of $p_0$ has to be chosen as negative. For $u^0<0$ the sign is taken to be
positive. However the absolute value of $u^0$ in \eqref{eq:zeroth-momentum-component} produces an additional minus
sign in this case. This leads to:
\begin{align}
\label{eq:classical-lagrangian}
L&=-p_0u^0-\mathbf{p}\cdot \mathbf{u}=\frac{\Upsilon m_{\psi}(u^0)^2}{\sqrt{\Upsilon^2(u^0)^2-\mathbf{u}^2}}-\frac{m_{\psi}\mathbf{u}^2}{\Upsilon\sqrt{\Upsilon^2(u^0)^2-\mathbf{u}^2}}=m_{\psi}\sqrt{(u^0)^2-\frac{\mathbf{u}^2}{\Upsilon^2}} \notag \\
&=m_{\psi}\sqrt{(u\cdot\xi)^2-\frac{1}{\Upsilon^2}\left[(u\cdot\xi)^2-u^2\right]}\,,
\end{align}
with the preferred timelike direction $(\xi^{\mu})=(1,0,0,0)^T$.
If only the positive-energy solution in \eqref{eq:zeroth-momentum-component} is considered, the Lagrangian with a global minus sign must be
taken into account as well. It can be checked that Eqs.~(\ref{eq:modified-generic-dispersion-law}) --
(\ref{eq:group-velocities}) are fulfilled by the positive Lagrangian for $u^0<0$ and by the negative Lagrangian for $u^0\geq 0$. Since
in the remainder of the paper $u^0\geq 0$ will be chosen anyhow, the Lagrangian with a global minus sign will be considered from now on.
For $\Upsilon=1$ one obtains the standard result $L=\pm m_{\psi}\sqrt{u_{\mu}u^{\mu}}$. The Lagrangian itself has an intrinsic metric
$r_{\mu\nu}$ associated to it, which is used to define scalar products, e.g., $u\cdot \xi=r_{\mu\nu}u^{\mu}\xi^{\nu}$. This intrinsic metric
corresponds to the Minkowski metric, i.e., $r_{\mu\nu}=\eta_{\mu\nu}$.

The following section, which ought to be understood as an interlude, is dedicated to identifying the Finsler structure associated to the Lagrangian
with a global minus sign, i.e., $L=-m_{\psi}\sqrt{(u^0)^2-\mathbf{u}^2/\Upsilon^2}$ (see \cite{Bao:2000} for the properties of such a structure).
As outlined in the introduction, the basic goal of the community is to understand how Lorentz-violating theories can be coupled to gravitational
backgrounds in a consistent manner. A reasonable assumption is that this is possible using the concept of Finsler geometry, since the latter
incorporates intrinsic preferred directions into the description of geometrical quantities in a natural way. Readers who are not interested in
Finsler geometry can skip the rest of the current section, since the results will not be directly employed in the remainder of the article.

There are two different possibilities of proceeding \cite{Kostelecky:2011qz}.
The first is to set $u^0=0$, which results in a three-dimensional Finsler structure describing a Euclidean geometry with a global scaling factor:
\begin{equation}
\widetilde{F}_{\Upsilon}(y)\equiv \frac{\mathrm{i}}{m_{\psi}}L(u^0=0,u^i=y^i)=\frac{1}{\Upsilon}\sqrt{r_{ij}y^iy^j}\,,\quad (r_{ij})=\mathrm{diag}(1,1,1)\,,\quad y\in TM\setminus \{0\}\,,
\end{equation}
where $TM$ is the tangent bundle of the Finsler space. The scalar product of two vectors $\alpha$, $\beta$ in the tangent space is given by
$\alpha\cdot \beta=r_{ij}\alpha^i\beta^j$ with the intrinsic metric $(r_{ij})=\mathrm{diag}(1,1,1)$. This structure describes a Euclidean space with
all dimensions scaled by $1/\Upsilon$. A similar space results from applying the same procedure to the Finsler structure of the
nonminimal coefficient $m^{(5)00}$ considered in \cite{Schreck:2014hga}.

The alternative is to perform a Wick rotation leading to the four-dimensional Finsler structure
\begin{align}
\label{eq:finsler-structure}
F_{\Upsilon}(y)&\equiv \frac{\mathrm{i}}{m_{\psi}}L(u^0=\mathrm{i}y^4,u^i=y^i)=\sqrt{(y^4)^2+\frac{1}{\Upsilon^2}\sum_{i=1,2,3} (y^i)^2}=\sqrt{(y\cdot \zeta)^2+\frac{1}{\Upsilon^2}\left[y^2-(y\cdot \zeta)^2\right]}\,,
\end{align}
where $y\in TM\setminus \{0\}$. The intrinsic metric here is $(r_{ij})=\mathrm{diag}(1,1,1,1)$ and $\zeta=(0,0,0,1)^T$ is a preferred direction
where $\zeta^i\equiv \xi^i$ for $i=1\dots 3$ and $\zeta^4\equiv\xi^0$ with the $\xi^{\mu}$ used in \eqref{eq:classical-lagrangian}.
The following considerations will be concentrated on the second Finsler structure $F_{\Upsilon}$. The Finsler metric can be computed via%%
\begin{equation}
\label{eq:finsler-metric}
g_{ij}(y)\equiv \frac{1}{2}\frac{\partial}{\partial y^i}\frac{\partial}{\partial y^j}F_{\Upsilon}(y)^2\,,\quad
(g_{ij})=\mathrm{diag}\left(\frac{1}{\Upsilon^2},\frac{1}{\Upsilon^2},\frac{1}{\Upsilon^2},1\right)\,,
\end{equation}
and the particular result is independent of $y$.
The Finsler structure $F_{\Upsilon}$ describes a Euclidean geometry as well. To check this, the Cartan torsion \cite{Bao:2004}
\begin{equation}
\label{eq:cartan-torsion}
C_{ijk}\equiv \frac{1}{2}\frac{\partial g_{ij}}{\partial y^k}=\frac{1}{4}\frac{\partial^3}{\partial y^i\partial y^j\partial y^k}F_{\Upsilon}^2\,.
\end{equation}
is needed where its mean is defined as
\begin{equation}
\label{eq:mean-cartan-torsion}
\mathbf{I}\equiv I_iy^i\,,\quad I_i\equiv g^{jk}C_{ijk}\,,\quad (g^{ij})\equiv (g_{ij})^{-1}\,,
\end{equation}
with the inverse Finsler metric $g^{ij}$. For the special Finsler metric in \eqref{eq:finsler-metric} the mean Cartan torsion $\mathbf{I}$
vanishes, which according to Deicke's theorem \cite{Deicke:1953} shows that the corresponding space is Riemannian.\footnote{In
\cite{Kostelecky:2011qz} Lagrangians were considered with their intrinsic metric $r_{\mu\nu}$ being promoted to a general pseudo-Riemannian
metric. By doing so, the Lagrangian can describe the motion of a relativistic particle on a curved spacetime manifold. Performing the
generalization here would lead to the Finsler structure of \eqref{eq:finsler-structure} with its scalar products being defined by
an intrinsic metric $r_{ij}$, which is not necessarily flat. In this case according to \eqref{eq:finsler-metric} the Finsler metric 
$g_{ab}=r_{aj}r_{bm}\zeta^j\zeta^m+(r_{ab}-r_{aj}r_{bm}\zeta^j\zeta^m)/\Upsilon^2$ would be associated to the structure. Note
that $\Upsilon$, $\zeta^a$, and $r_{ab}$ are then understood to be position-dependent functions, in general.
Since $g_{ab}$ does not depend on $y^i$, its mean Cartan torsion vanishes showing that it still describes a Riemannian
space. In the remainder of the current article the intrinsic metric will be assumed to be flat, though.} In this space three
dimensions are scaled by $1/\Upsilon$ and one dimension remains standard. Therefore the length of a vector in the scaled subspace, which corresponds to
the spatial part of the original spacetime, is scaled where the angle between such vectors stays unmodified. However angles between
vectors change when they have one component pointing along the $y^4$-axis, which has some influence on, e.g., velocities in the corresponding
spacetime.

All Finsler spaces in the context of the minimal SME, which have been considered in other references so far, are related to non-Euclidean
spaces. This holds for a-space \cite{Kostelecky:2010hs,Kostelecky:2011qz}, b-space \cite{Kostelecky:2010hs,Kostelecky:2011qz}, the
bipartite spaces \cite{Kostelecky:2012ac}, and the spaces considered in \cite{Colladay:2012rv}. A reasonable conjecture is that only
single, isotropic dispersion relations such as the one investigated here lead to Euclidean structures.

%---------------------------------------------------------------------------------------------------
\section{Charged relativistic particle in an electromagnetic field}
\label{sec:particle-electromagnetic-field}
%---------------------------------------------------------------------------------------------------

After clarifying the mathematical foundations of the modified Lagrangian in the last section, its physical properties shall be investigated. In
what follows, particle trajectories shall be parameterized such that $u^0=c$ and $\mathbf{u}=\mathbf{v}$ where $c$ is the speed of light and
$\mathbf{v}$ the ordinary three-velocity of the particle. Note that natural coordinates are used with $c=1$. The Lagrangian then reads as
follows:
\begin{equation}
L=-m_{\psi}\sqrt{1-\frac{\mathbf{v}^2}{\Upsilon^2}}\,.
\end{equation}
If the particle moves freely, the trajectory will be the same straight line such as in the standard case without any Lorentz violation. Using the
metric corresponding to the Lagrangian,
\begin{equation}
g_{\mu\nu}=\frac{1}{2}\frac{\partial^2L^2}{\partial u^{\mu}\partial u^{\nu}}=m_{\psi}^2\,\mathrm{diag}\left(1,-\frac{1}{\Upsilon^2},-\frac{1}{\Upsilon^2},-\frac{1}{\Upsilon^2}\right)_{\mu\nu}\,,
\end{equation}
the conserved quantities can be computed according to Eqs. (35) and (36) in \cite{Girelli:2006fw} and they are given by:
\begin{subequations}
\begin{align}
\mathcal{E}&=-\frac{1}{L}g_{0\nu}u^{\nu}=\frac{m_{\psi}}{\sqrt{1-\mathbf{v}^2/\Upsilon^2}}=\gamma_{\scriptscriptstyle{\Upsilon}}m_{\psi}\,, \\[2ex]
\mathcal{P}_i&=-\frac{1}{L}g_{i\nu}u^{\nu}=-\frac{m_{\psi}}{\sqrt{1-\mathbf{v}^2/\Upsilon^2}}\frac{v^i}{\Upsilon^2}=-\frac{\gamma_{\scriptscriptstyle{\Upsilon}}m_{\psi} v^i}{\Upsilon^2}\,,\quad \gamma_{\scriptscriptstyle{\Upsilon}}\equiv\frac{1}{\sqrt{1-\mathbf{v}^2/\Upsilon^2}}\,,
\end{align}
\end{subequations}
with a modified Lorentz factor $\gamma_{\scriptscriptstyle{\Upsilon}}$. 
For $\Upsilon=1$ these correspond to the classical energy and spatial momentum (besides a global sign), as expected. The spatial
momentum is part of the contravariant four-momentum, whereby the index on $\mathcal{P}_i$ has to be raised to produce an additional sign.
The quantities $\mathcal{E}$ and $\mathcal{P}_i$ will appear again below.

To understand
the modified physics, the classical particle is assigned an electric charge $q$ and its propagation in an electromagnetic field will be
studied. Therefore a four-potential $(A^{\mu})=(\phi,\mathbf{A})$ is introduced and the charged, classical particle is described by the
following Lagrangian:
\begin{equation}
\label{eq:lagrangian-electromagnetic-field}
L_{\mathrm{em}}=L+q\:\!\mathbf{v}\cdot\mathbf{A}-q\phi=-m_{\psi}\sqrt{1-\frac{\mathbf{v}^2}{\Upsilon^2}}+q\:\!\mathbf{v}\cdot\mathbf{A}-q\phi\,,
\end{equation}
with the scalar potential $\phi$ and the vector potential $\mathbf{A}$.  The equations of motion are obtained from the Euler-Lagrange
equations (with the position vector $\mathbf{x}$), which for the particular Lagrangian of \eqref{eq:lagrangian-electromagnetic-field} read
as follows:
\begin{subequations}
\begin{align}
\frac{\mathrm{d}}{\mathrm{d}t}\frac{\partial L_{\mathrm{em}}}{\partial\mathbf{v}}&=\frac{\partial L_{\mathrm{em}}}{\partial\mathbf{x}}\,, \\[2ex]
\label{eq:eom-electromagnetic-field}
\frac{\mathrm{d}}{\mathrm{d}t}\left(\frac{m_{\psi}\mathbf{v}/\Upsilon^2}{\sqrt{1-\mathbf{v}^2/\Upsilon^2}}+q\mathbf{A}\right)&=-q\boldsymbol{\nabla}\phi+q\boldsymbol{\nabla}(\mathbf{v}\cdot\mathbf{A})\,.
\end{align}
\end{subequations}
The total time derivative of the vector potential
\begin{equation}
\frac{\mathrm{d}\mathbf{A}}{\mathrm{d}t}=-\mathbf{v}\times (\boldsymbol{\nabla}\times \mathbf{A})+\boldsymbol{\nabla}(\mathbf{v}\cdot\mathbf{A})+\frac{\partial\mathbf{A}}{\partial t}\,,
\end{equation}
is used to express the right-hand side of \eqref{eq:eom-electromagnetic-field} via the physical electric and magnetic fields $\mathbf{E}$,
$\mathbf{B}$:%%
\begin{equation}
\label{eq:eom-spatial}
\frac{\mathrm{d}\mathbf{p}}{\mathrm{d}t}=q\mathbf{v}\times (\boldsymbol{\nabla}\times \mathbf{A})+q\left(-\boldsymbol{\nabla}\phi-\frac{\partial \mathbf{A}}{\partial t}\right)=q\:\!\mathbf{v}\times \mathbf{B}+q\mathbf{E}\,.
\end{equation}
For the zeroth four-momentum component, i.e., the particle energy, a further equation can be derived directly from the equations of motion
for the spatial momentum components:
\begin{equation}
\label{eq:eom-zeroth}
\frac{\mathrm{d}p_0}{\mathrm{d}t}=\frac{\Upsilon^2}{p_0}\mathbf{p}\cdot \frac{\mathrm{d}\mathbf{p}}{\mathrm{d}t}=\frac{1}{\gamma_{\scriptscriptstyle{\Upsilon}}m_{\psi}}\gamma_{\scriptscriptstyle{\Upsilon}}m_{\psi}\mathbf{v}\cdot \frac{\mathrm{d}\mathbf{p}}{\mathrm{d}t}=q\:\!\mathbf{v}\cdot\mathbf{E}\,.
\end{equation}
Now a relativistic momentum and energy can be introduced via
\begin{equation}
\label{eq:relativistic-momentum}
\mathbf{p}=\frac{\gamma_{\scriptscriptstyle{\Upsilon}}m_{\psi}\mathbf{v}}{\Upsilon^2}\,,\quad p_0=\gamma_{\scriptscriptstyle{\Upsilon}}m_{\psi}\,,\quad \gamma_{\scriptscriptstyle{\Upsilon}}=\frac{1}{\sqrt{1-\mathbf{v}^2/\Upsilon^2}}\,.
\end{equation}
In fact, with the free Lagrangian $L$ it can be cross-checked that
\begin{equation}
p_0=-\frac{\partial L}{\partial u^0}=m_{\psi}\frac{u^0}{\sqrt{(u^0)^2-\mathbf{u}^2/\Upsilon^2}}\,,\quad (p_i)=-\frac{\partial L}{\partial\mathbf{u}}=-m_{\psi}\frac{\mathbf{u}/\Upsilon^2}{\sqrt{(u^0)^2-\mathbf{u}^2/\Upsilon^2}}\,,
\end{equation}
which corresponds to \eqref{eq:relativistic-momentum} for $u^0=c=1$, $\mathbf{u}=\mathbf{v}$ and again taking into account that $\mathbf{p}$
is the spatial momentum of the contravariant momentum four-vector. Therefore raising the index on $p_i$ produces an additional minus sign.
With the proper time $\mathrm{d}\tau=\mathrm{d}t/\gamma$
the equations of motion (\ref{eq:eom-spatial}), (\ref{eq:eom-zeroth}) can be written in a covariant form:
\begin{equation}
\label{eq:modified-eom}
\frac{\mathrm{d}\widetilde{u}^{\alpha}}{\mathrm{d}\tau}=\frac{q}{m_{\psi}}F^{\alpha\beta}u_{\beta}\,,\quad (\widetilde{u}^{\alpha})=\gamma_{\scriptscriptstyle{\Upsilon}}\begin{pmatrix}
1 \\
\mathbf{v}/\Upsilon^2 \\
\end{pmatrix}\,,\quad (u^{\alpha})=\gamma\begin{pmatrix}
1 \\
\mathbf{v} \\
\end{pmatrix}\,,
\end{equation}
where $F_{\mu\nu}=\partial_{\mu}A_{\nu}-\partial_{\nu}A_{\mu}$ is the electromagnetic field strength tensor.
Note that the four-velocity $\widetilde{u}^{\alpha}$ used on the left-hand side of the latter equation involves both modifications in the Lorentz factor and
the spatial velocity components, whereas the four-velocity $u^{\alpha}$ on the right-hand side is standard. The reason for this
is that the particle kinematics is modified by the Lorentz-violating background field in contrast to its coupling to the electromagnetic field.

Now the modified equations of motion shall be solved for particular cases to understand how their solutions are affected by Lorentz violation. First, consider
the case of a vanishing electric field, $\mathbf{E}=\mathbf{0}$, where the particle moves perpendicularly to a magnetic field $\mathbf{B}=B\,\!\widehat{\mathbf{e}}_z$,
i.e., its initial velocity and position shall be given by $\mathbf{v}(0)=v\,\!\widehat{\mathbf{e}}_y$ and $\mathbf{x}(0)=R\:\!\widehat{\mathbf{e}}_x$, respectively.
Here $v$ is the constant velocity and $R$ the particle distance from the origin at the beginning.
The equations of motion in the laboratory frame read
\begin{equation}
\frac{\mathrm{d}}{\mathrm{d}t}\left(\frac{m_{\psi}\mathbf{v}/\Upsilon^2}{\sqrt{1-\mathbf{v}^2/\Upsilon^2}}\right)=q\;\!\mathbf{v}\times \mathbf{B} \Leftrightarrow \frac{\gamma_{\scriptscriptstyle{\Upsilon}}m_{\psi}}{\Upsilon^2}\frac{\mathrm{d}\mathbf{v}}{\mathrm{d}t}=q\;\!\mathbf{v}\times\mathbf{B}\,,
\end{equation}
where $\gamma_{\scriptscriptstyle{\Upsilon}}$ is time-independent, since the magnitude of the velocity does not change in a magnetic field. The latter
differential equations with the initial conditions above are satisfied by the following time-dependent particle position and velocity:
\begin{subequations}
\label{eq:circular-movement-magnetic-field-complete}
\begin{align}
\label{eq:circular-movement-magnetic-field}
\mathbf{x}(t)&=\begin{pmatrix}
R\cos(\omega t) \\
R\sin(\omega t) \\
0 \\
\end{pmatrix}\,,\quad \mathbf{v}(t)=R\omega\begin{pmatrix}
-\sin(\omega t) \\
\cos(\omega t) \\
0 \\
\end{pmatrix}\,, \\[2ex]
v&=R|\omega|\,,\quad \omega=-\frac{\Upsilon^2C}{\sqrt{1+\Upsilon^2C^2R^2}}=-\Upsilon^2C\left(1-\frac{1}{2}\Upsilon^2C^2R^2+\dots\right)\,,\quad C=\frac{qB}{m_{\psi}}\,.
\end{align}
\end{subequations}
This describes a circular motion with radius $R$ and angular frequency $\omega$ such as in the standard case where the sign gives the rotational
direction. However additional scaling factors $\Upsilon$ appear that can be explained as follows. Kinematic quantities
living in tangent space such as $R$ and $v$ involve a length scale and, therefore, get multiplied by one power of $1/\Upsilon$ each (cf.~\eqref{eq:finsler-metric}):
$\{R,v\}\mapsto (1/\Upsilon)\{R,v\}$.
In contrast to position and velocity vectors, both the momentum vector $\mathbf{p}$ and the vector potential $\mathbf{A}$ live in cotangent space and
have an inverse length scale associated to them, which is why they are multiplied by $\Upsilon$ (see the dispersion relation of
\eqref{eq:modified-generic-dispersion-law}): $\{\mathbf{p},\mathbf{A}\}\mapsto \Upsilon\{\mathbf{p},\mathbf{A}\}$. Taking into account the relationship
$B^l=-\mathrm{i}\varepsilon^{ijk}p_jA^k$ in momentum space, it follows that $\mathbf{B}\mapsto \Upsilon^2\mathbf{B}$. Because of $v=R|\omega|$ the
angular frequency $\omega$ is unaffected by the scaling. Concerning the particle motion, since the magnitude of the velocity is constant, $p_0$
is time-independent. Due to $\mathbf{E}=\mathbf{0}$ the first of \eqref{eq:modified-eom} is satisfied as well, for consistency.

As a next example consider a particle moving in a vanishing magnetic field, $\mathbf{B}=\mathbf{0}$, where the particle initially moves perpendicularly
to the electric field $\mathbf{E}=E\:\!\widehat{\mathbf{e}}_z$ with the initial conditions $\mathbf{v}(0)=v_0\widehat{\mathbf{e}}_y$ and
$\mathbf{x}(0)=\mathbf{0}$. The equations of motion in the laboratory frame then read as follows:
\begin{equation}
\frac{\mathrm{d}}{\mathrm{d}t}\left(\frac{m_{\psi}\mathbf{v}/\Upsilon^2}{\sqrt{1-\mathbf{v}^2/\Upsilon^2}}\right)=q\mathbf{E}\,.
\end{equation}
An integration with respect to $t$ using the initial condition $\mathbf{v}(0)=v_0\widehat{\mathbf{e}}_y$ leads to:
\begin{equation}
\frac{\mathbf{v}}{\sqrt{1-\mathbf{v}^2/\Upsilon^2}}-\frac{v_0\widehat{\mathbf{e}}_y}{\sqrt{1-v_0^2/\Upsilon^2}}=\frac{q\Upsilon^2}{m_{\psi}}\mathbf{E}t\,.
\end{equation}
This is a system of algebraic equations for $v_y$ and $v_z$ ($v_x\equiv 0$), which can be solved to give
\begin{subequations}
\begin{equation}
\label{eq:velocities-e-field-problem}
v_y(t)=\frac{v_0}{\sqrt{1+\Upsilon^2\widetilde{C}^2t^2}}\,,\quad v_z(t)=\frac{\Upsilon^2\widetilde{C}t}{\sqrt{1+\Upsilon^2\widetilde{C}^2t^2}}\,,\quad \widetilde{C}=\frac{qE}{m_{\psi}}\sqrt{1-\frac{v_0^2}{\Upsilon^2}}\,,
\end{equation}
where a subsequent integration results in
\begin{equation}
y(t)=\frac{v_0}{\Upsilon\widetilde{C}}\ln\left(\Upsilon\widetilde{C} t+\sqrt{1+\Upsilon^2\widetilde{C}^2t^2}\,\right)\,,\quad z(t)=\frac{\Upsilon^2\widetilde{C}t^2}{1+\sqrt{1+\Upsilon^2\widetilde{C}^2t^2}}\,.
\end{equation}
\end{subequations}
Here the trajectory again involves additional scalings with $\Upsilon$. The behavior can be understood when taking into account that each of the velocities
$v_y$, $v_z$, $v_0$ and positions $y$, $z$ gets one power of $1/\Upsilon$: $\{v_y,v_z,v_0,y,z\}\mapsto (1/\Upsilon)\{v_y,v_z,v_0,y,z\}$. Due to $E^j=\mathrm{i}(p_j\phi+p_0A^j)$ and the
scaling $\{\mathbf{p},\mathbf{A}\}\mapsto \Upsilon\{\mathbf{p},\mathbf{A}\}$, the electric field is subject to $\mathbf{E}\mapsto \Upsilon\mathbf{E}$.
Concerning the physical behavior of the particle, the velocity component in
$y$-direction goes to zero starting from its initial value $v_0$. This is due to the relativistic increase in mass, since there is no force along the
$y$-direction compensating for this effect. Therefore the distance traveled in $y$-direction grows logarithmically, i.e., very slowly.  The velocity in
$z$-direction steadily increases to reach its maximum value $v_z(t=\infty)=\Upsilon$ as expected. For large times the particle then travels with the
practically constant velocity $\Upsilon$ resulting in the uniform motion $z(t)\simeq \Upsilon t$ for $t\mapsto \infty$. For consistency, the first of
\eqref{eq:modified-eom} is fulfilled when inserting the electric field vector and the velocity components
of \eqref{eq:velocities-e-field-problem}.

\subsection{Introduction of particle spin}

Since spin is a manifestly quantum-mechanical concept, the classical particle studied in the previous sections does not have any spin
associated to it, although it shall be based on a Lorentz-violating fermion. However it is possible to introduce
spin for a classical particle according to the lines of \cite{Bargmann:1959gz}. The authors of the latter reference derive a relativistic
equation of motion (often denoted as the BMT equation where this abbreviation refers to the authors' second names) for the spin of a classical particle of
electric charge $q$ and mass $m_{\psi}$ in an electromagnetic field. By doing so, they take the equation of motion
for the spin three-vector $\mathbf{s}$ as a basis:%%
\begin{equation}
\label{eq:spin-eom-rest-frame}
\frac{\mathrm{d}\mathbf{s}}{\mathrm{d}\tau}=\frac{gq}{2m_{\psi}}(\mathbf{s}\times \mathbf{B})\,.
\end{equation}
Generalization to arbitrary frames leads to the BMT equation:
\begin{equation}
\label{eq:spin-eom}
\frac{\mathrm{d}s^{\alpha}}{\mathrm{d}\tau}=\frac{gq}{2m_{\psi}}\left[F^{\alpha\beta}s_{\beta}+(F^{\beta\gamma}s_{\beta}u_{\gamma})u^{\alpha}\right]-\left(\frac{\mathrm{d}u^{\beta}}{\mathrm{d}\tau}s_{\beta}\right)u^{\alpha}\,.
\end{equation}
Here $g$ is the Land\'{e} factor of the particle, $u^{\mu}$ is the particle velocity, $(s^{\mu})=(s^0,\mathbf{s})$ the spin
four-vector, and $\tau$ the proper time.
The spin four-vector can be understood as a covariant particle polarization vector. According to \cite{Bargmann:1959gz} it is the
expectation value of the Pauli-Lubanski (pseudo)vector $W^{\mu}\equiv \varepsilon^{\mu\nu\varrho\sigma}M_{\nu\varrho}p_{\sigma}/2$, which
is a four-vector by construction \cite{Lubanski:1942}.\footnote{The operator stated on the first page of \cite{Lubanski:1942} is $-W^2p^2$ with
the metric used here.} Here $\varepsilon^{\mu\nu\varrho\sigma}$ is the four-dimensional Levi-Civita symbol with the condition $\varepsilon^{0123}=1$
and $M_{\nu\varrho}$ are the generators of the Lorentz group. The latter are a covariant generalization 
of the angular momentum three-vector. The scalar product of $W^{\mu}$ with itself is one of the two Casimir operators of the Poincar\'{e} algebra
giving the eigenvalues $W^2=-m_{\psi}s(s+1)$ for a particle at rest. Since $s$ is the particle spin eigenvalue, $W^{\mu}$ can be
interpreted as a covariant generalization of the nonrelativistic spin operator.

The quantum mechanical treatment of fermion spin in the SME was carried out in \cite{Kostelecky:2013rta,Gomes:2014kaa}.
In the latter references the time evolution of the spin expectation value was obtained from the expectation value of the commutator of the spin
operator and the Lorentz-violating Hamiltonian  (cf.~\cite{Bluhm:1999dx} where this approach was introduced). At first order in Lorentz violation
a Larmor-like precession of the particle spin is induced by
controlling coefficients leading to two distinct fermion dispersion relations. These are subsets of the effective $\widehat{g}$ and $\widehat{H}$
operators that comprise the operators $\widehat{b}^{\mu}$, $\widehat{d}^{\mu}$, $\widehat{H}^{\mu\nu}$, and $\widehat{g}^{\mu\nu}$. Such
a behavior is reminiscent of the standard case when spin precession occurs for the valence electron of a hydrogen atom in an external magnetic
field accompanied by a splitting of its energy levels.

Therefore considering dispersion laws of
the form of \eqref{eq:modified-generic-dispersion-law}, the only set of isotropic controlling coefficients that may lead to spin precession is given
by \eqref{eq:second-set-isotropic-g-coefficients}. However it is evident that their correction to the standard fermion dispersion relation
is of quadratic order. Inserting these coefficients into Eq.~(62) of \cite{Kostelecky:2013rta}, which controls the spin part of the Hamiltonian at
first order in Lorentz violation, gives zero as expected. Therefore the spin operator commutes with the Hamiltonian at first order in the controlling
coefficients for the sets of coefficients studied. Based on the Heisenberg equations no additional time dependence of the spin operator emerges
from Lorentz violation at first order for the sets of coefficients leading to the classical Lagrange function of \eqref{eq:classical-lagrangian}. Note that the set of
coefficients given by \eqref{eq:first-set-isotropic-g-coefficients} results in two distinct isotropic dispersion relations and classical Lagrangians corresponding
to operators with these properties are not considered in the current article.

To ensure that Lorentz violation does not give rise to an additional contribution
to the zeroth component of the BMT equation, the explicit form of the Pauli-Lubanski vector can be examined. In matrix form the
generators of the Lorentz group are written as
\begin{equation}
(M_{\mu\nu})=\begin{pmatrix}
0 & -K_1 & -K_2 & -K_3 \\
K_1 & 0 & J_3 & -J_2 \\
K_2 & -J_3 & 0 & J_1 \\
K_3 & J_2 & -J_1 & 0 \\
\end{pmatrix}\,,
\end{equation}
where $K_i$ and $J_i$ (for $i=1\dots 3$) are the boost and rotation generators, respectively. Now the Pauli-Lubanski vector can be
computed to give explicitly
\begin{equation}
(W^{\mu})=-\begin{pmatrix}
\mathbf{J}\cdot \mathbf{p} \\
p^0J_1+K_2p^3-K_3p^2 \\
p^0J_2+K_3p^1-K_1p^3 \\
p^0J_3+K_1p^2-K_2p^1 \\
\end{pmatrix}\,.
\end{equation}
where $\mathbf{J}=(J_1,J_2,J_3)$. Replacing $\mathbf{J}$ by the spin operator $-\boldsymbol{\sigma}/2$ where $\boldsymbol{\sigma}=(\sigma^1,\sigma^2,\sigma^3)$
with the three Pauli matrices $\sigma^i$ ($i=1\dots 3$) we see that $W^0=\boldsymbol{\sigma}\cdot \mathbf{p}/2$. Since the spin
part of the Lorentz-violating Hamiltonian is of second order in Lorentz violation for the coefficients considered, $W^0$ commutes
with the Hamiltonian at first order in Lorentz violation. Thus an additional time-dependence for the zeroth component of the BMT
equation that is caused by Lorentz violation is not expected at first order. For these reasons the validity of the BMT equation, as it stands in
\eqref{eq:spin-eom}, is granted at first order in Lorentz violation for the particular isotropic frameworks considered. In a few lines below an additional
statement will be made about possible Lorentz-violating effects at second order.

However Lorentz violation may still have an influence on particle spin at first order due to the second term on the right-hand side of
\eqref{eq:spin-eom}, which is linked to particle kinematics. It involves the four-acceleration, which allows us to use the particle
equations of motion. Now the isotropic Lorentz-violating coefficients are assumed to be much smaller than one, i.e, $\Upsilon=1+\chi$
with a generic, dimensionless, isotropic Lorentz-violating coefficient $\chi$. The modified four-velocity $\widetilde{u}^{\alpha}$ of the
particle, which has been employed on the left-hand side of \eqref{eq:modified-eom}, is then expanded around a zero coefficient $\chi$:
\begin{equation}
(\widetilde{u}^{\alpha})=\gamma\begin{pmatrix}
1 \\
\mathbf{v} \\
\end{pmatrix}-\gamma^3\begin{pmatrix}
v^2 \\
(2-v^2)\mathbf{v} \\
\end{pmatrix}\chi\,,
\end{equation}
with the magnitude $v\equiv |\mathbf{v}|$ of the three-velocity $\mathbf{v}$.
Therefore for a small Lorentz-violating coefficient $\chi$ the equations of motion of the classical particle involve the standard terms
plus an additional contribution on the right-hand side, which is linear in $\chi$:
\begin{equation}
\frac{\mathrm{d}u^{\alpha}}{\mathrm{d}\tau}=\frac{q}{m_{\psi}}F^{\alpha\beta}u_{\beta}+\frac{\mathrm{d}}{\mathrm{d}\tau}\left[\gamma^3\begin{pmatrix}
v^2 \\
(2-v^2)\mathbf{v} \\
\end{pmatrix}^{\alpha}\,
\right]\chi\,.
\end{equation}
The nonrelativistic version of this equation is obtained by expanding all quantities with respect to $v^2\ll 1$:%%
\begin{equation}
\label{eq:eom-nonrelativistic}
\frac{\mathrm{d}}{\mathrm{d}t}\begin{pmatrix}
1+v^2/2 \\
\mathbf{v}
\end{pmatrix}^{\alpha}=\frac{q}{m_{\psi}}F^{\alpha\beta}\begin{pmatrix}
1 \\
-\mathbf{v} \\
\end{pmatrix}_{\beta}+\frac{\mathrm{d}}{\mathrm{d}t}\begin{pmatrix}
v^2 \\
2\mathbf{v} \\
\end{pmatrix}^{\alpha}\chi\,.
\end{equation}
This is a set of four nonrelativistic equations where the first one gives a relation for the nonrelativistic kinetic energy of the particle
in the electromagnetic field and the remaining ones give the acceleration caused by the Lorentz force. For $\chi\ll 1$ they are given by
\begin{subequations}
\begin{align}
\frac{\mathrm{d}}{\mathrm{d}t}\frac{v^2}{2}&=(1+2\chi)\frac{q}{m_{\psi}}\mathbf{E}\cdot\mathbf{v}\,, \\[2ex]
\label{eq:eom-particle-velocity}
\frac{\mathrm{d}\mathbf{v}}{\mathrm{d}t}&=(1+2\chi)\frac{q}{m_{\psi}}(\mathbf{E}+\mathbf{v}\times \mathbf{B})\,,
\end{align}
\end{subequations}
and they will be needed shortly. 
Then the modified evolution equations for the particle spin can be obtained by inserting \eqref{eq:eom-nonrelativistic} in \eqref{eq:spin-eom},
neglecting all contributions of the order of $v^2\ll 1$, and assuming $|v\dot{v}|\ll |\dot{\mathbf{v}}|$:
\begin{align}
\frac{\mathrm{d}s^{\alpha}}{\mathrm{d}t}&=\frac{gq}{2m_{\psi}}\left[F^{\alpha\beta}s_{\beta}+(F^{\beta\gamma}s_{\beta}u_{\gamma})u^{\alpha}\right]-\frac{q}{m_{\psi}}(F^{\beta\gamma}s_{\beta}u_{\gamma})u^{\alpha}-\begin{pmatrix}
0 \\
2\dot{\mathbf{v}}\chi \\
\end{pmatrix}^{\beta}s_{\beta}u^{\alpha} \notag \\
&=\frac{q}{m_{\psi}}\left[\frac{g}{2}F^{\alpha\beta}s_{\beta}+\left(\frac{g}{2}-1\right)(F^{\beta\gamma}s_{\beta}u_{\gamma})u^{\alpha}\right]+2\chi\dot{\mathbf{v}}\cdot \mathbf{s}\,u^{\alpha}\,.
\end{align}
The intermediate result is that there appears an additional term on the right-hand side of the spin evolution equations, which is
proportional to the Lorentz-violating coefficient and describes a coupling between the spin vector and the ordinary particle
three-acceleration $\dot{\mathbf{v}}$. Introducing the electromagnetic fields leads to
\begin{subequations}
\begin{align}
\frac{\mathrm{d}}{\mathrm{d}t}\begin{pmatrix}
s^0 \\
\mathbf{s} \\
\end{pmatrix}&=\frac{q}{m_{\psi}}\left\{\frac{g}{2}\begin{pmatrix}
\mathbf{E}\cdot\mathbf{s} \\
\mathbf{E}s^0+\mathbf{s}\times\mathbf{B} \\
\end{pmatrix}\right.
\left.+\left(\frac{g}{2}-1\right)\left[(\mathbf{E}\cdot \mathbf{v})s^0-(\mathbf{E}+\mathbf{v}\times \mathbf{B})\cdot \mathbf{s}+\frac{m_{\psi}}{q}\widetilde{\chi}\,\dot{\mathbf{v}}\cdot\mathbf{s}\right]\begin{pmatrix}
1 \\
\mathbf{v} \\
\end{pmatrix}\right\}\,, \\
\widetilde{\chi}&\equiv\frac{4\chi}{g-2}\,,
\end{align}
\end{subequations}
where the coefficient $\widetilde{\chi}$ has been introduced for convenience.
Using the equations of motion (\ref{eq:eom-particle-velocity}), the Lorentz-violating contribution can be combined with
the coupling term between the electromagnetic fields and the spatial spin vector:
\begin{equation}
-(\mathbf{E}+\mathbf{v}\times \mathbf{B})\cdot \mathbf{s}+\frac{m_{\psi}}{q}\widetilde{\chi}\,\dot{\mathbf{v}}\cdot\mathbf{s}=-\big(1-\widetilde{\chi}\big)[\mathbf{E}+\mathbf{v}\times\mathbf{B}]\cdot\mathbf{s}\,.
\end{equation}
For a vanishing electric field, $\mathbf{E}=\mathbf{0}$, the spin evolution equations then give
\begin{equation}
\label{eq:spin-eom-nonrelativistic}
\frac{\mathrm{d}}{\mathrm{d}t}\begin{pmatrix}
s^0 \\
\mathbf{s} \\
\end{pmatrix}=\frac{q}{m_{\psi}}\left\{\frac{g}{2}\begin{pmatrix}
0 \\
\mathbf{s}\times\mathbf{B} \\
\end{pmatrix}+\left(1-\frac{g}{2}\right)\big(1-\widetilde{\chi}\big)[\mathbf{v}\times\mathbf{B}]\cdot\mathbf{s} \begin{pmatrix}
1 \\
\mathbf{v} \\
\end{pmatrix}\right\}\,.
\end{equation}
This is the final form of the BMT equation that shall be solved.
To make a physical prediction, a particular \textit{Ansatz} for the spin four-vector is inserted, which was introduced in \cite{Bargmann:1959gz}:
\begin{subequations}
\begin{align}
\label{eq:spin-ansatz}
s^{\alpha}&=\sqrt{-s^2}(e_l^{\alpha}\cos\phi+e_t^{\alpha}\sin\phi)\,, \\[2ex]
e_l^{\alpha}&=e_l^{\alpha}(t)=\begin{pmatrix}
v(t) \\
\widehat{\mathbf{v}}(t) \\
\end{pmatrix}^{\alpha}\,,\quad e_t^{\alpha}=e_t^{\alpha}(t)=\begin{pmatrix}
0 \\
\widehat{\mathbf{n}}(t) \\
\end{pmatrix}^{\alpha}\,, \\[2ex]
\widehat{\mathbf{v}}(t)&\equiv \frac{\mathbf{v}(t)}{v(t)}\,,\quad |\widehat{\mathbf{n}}(t)|=1\,,\quad \phi=\phi(t)\,.
\end{align}
\end{subequations}
The \textit{Ansatz} is chosen such that $s^2=1$ and $s\cdot u=0$. Both $e_l^{\alpha}$ and $e_t^{\alpha}$ are normalized
four-vectors where the additional $\gamma$-factor before $e_l^{\alpha}$ has been omitted, since the nonrelativistic regime is considered.
The spatial part of the first vector is chosen to point along the particle velocity where the spatial part of the second vector is assumed to be perpendicular
to the velocity, i.e.,
$\widehat{\mathbf{n}}\cdot \widehat{\mathbf{v}}=0$. Hence $s^{\alpha}$ is decomposed into a longitudinal and a transverse part. Since $s^2$ is
assumed to be constant, a change of the angle $\phi$ describes how the longitudinal part is transformed into the transverse part and vice versa.
The \textit{Ansatz} contains the velocity vector $\mathbf{v}(t)$ that is a solution of the equations of motion. As an example consider
the case of a vanishing electric field $\mathbf{E}=\mathbf{0}$ and a constant magnetic field pointing along the $z$ direction: $\mathbf{B}=B\,\widehat{\mathbf{e}}_z$.
The charged particle carries out a circular motion as derived in \eqref{eq:circular-movement-magnetic-field-complete} where the following quantities are
understood to be valid at linear order in the Lorentz-violating coefficient $\chi$. Thereby the choices below have to be made for the
\textit{Ansatz} of \eqref{eq:spin-ansatz}:
\begin{subequations}
\begin{align}
\mathbf{v}(t)&=(1+2\chi)\widetilde{v}\,\widehat{\mathbf{v}}\,,\quad \widehat{\mathbf{v}}(t)=\begin{pmatrix}
-\sin(\omega t) \\
\cos(\omega t) \\
0 \\
\end{pmatrix}\,,\quad \widehat{\mathbf{n}}(t)=-\begin{pmatrix}
\cos(\omega t) \\
\sin(\omega t) \\
0 \\
\end{pmatrix}\,, \\[2ex]
\omega&=-(1+2\chi)\frac{\widetilde{v}}{R}\,,\quad \widetilde{v}=\frac{qBR}{m_{\psi}}\,.
\end{align}
\end{subequations}
Note that the Lorentz-violating coefficient has been pulled out of the magnitude of the three-velocity vector to make its appearance explicit. So $v(t)$ is
understood to be $|\mathbf{v}(t)|$ and $\widetilde{v}$ is the particle velocity for vanishing Lorentz violation.
Then the \textit{Ansatz} for the spin four-vector reads as follows:
\begin{equation}
\begin{pmatrix}
s^0 \\
\mathbf{s} \\
\end{pmatrix}=\sqrt{1-(1+2\chi)^2\,\widetilde{v}^2\cos^2(\Omega t)}\begin{pmatrix}
(1+2\chi)\widetilde{v}\cos(\Omega t) \\
-\sin[(\omega+\Omega)t] \\
\cos[(\omega+\Omega)t] \\
0 \\
\end{pmatrix}\,,\quad \Omega\equiv\dot{\phi}\,.
\end{equation}
The time derivative of the spin four-vector expanded for $\widetilde{v}^2\ll 1$ is given by
\begin{equation}
\frac{\mathrm{d}}{\mathrm{d}t}\begin{pmatrix}
s^0 \\
\mathbf{s} \\
\end{pmatrix}=\begin{pmatrix}
-(1+2\chi)\widetilde{v}\Omega\sin(\Omega t) \\
-(\omega+\Omega)\cos[(\omega+\Omega)t] \\
-(\omega+\Omega)\sin[(\omega+\Omega)t] \\
0 \\
\end{pmatrix}+\mathcal{O}(\widetilde{v}^2)\,,
\end{equation}
where the square root of the normalization leads to higher-order terms in the nonrelativistic expansion. Using the previous results
one obtains
\begin{subequations}
\begin{align}
\mathbf{s}\times \mathbf{B}&=\begin{pmatrix}
-\sin[(\omega+\Omega)t] \\
\cos[(\omega+\Omega)t] \\
0 \\
\end{pmatrix}\times \begin{pmatrix}
0 \\
0 \\
B \\
\end{pmatrix}=B\begin{pmatrix}
\cos[(\omega+\Omega)t] \\
\sin[(\omega+\Omega)t] \\
0 \\
\end{pmatrix}+\mathcal{O}(\widetilde{v}^2)\,, \\[2ex]
\mathbf{v}\times \mathbf{B}&=(1+2\chi)\widetilde{v}\begin{pmatrix}
-\sin(\omega t) \\
\cos(\omega t) \\
0 \\
\end{pmatrix}\times \begin{pmatrix}
0 \\
0 \\
B \\
\end{pmatrix}=(1+2\chi)\widetilde{v}B\begin{pmatrix}
\cos(\omega t) \\
\sin(\omega t) \\
0 \\
\end{pmatrix}\,, \\[2ex]
(\mathbf{v}\times \mathbf{B})\cdot \mathbf{s}&=-(1+2\chi)\widetilde{v}B\Bigl\{\sin[(\omega+\Omega)t]\cos(\omega t)-\cos[(\omega+\Omega)t]\sin(\omega t)\Bigr\}+\mathcal{O}(\widetilde{v}^2) \notag \\
&=-(1+2\chi)\widetilde{v}B\sin(\Omega t)+\mathcal{O}(\widetilde{v}^2)\,.
\end{align}
\end{subequations}
Hence the right-hand side of the nonrelativistic BMT equation (\ref{eq:spin-eom-nonrelativistic}) reads
\begin{align}
\frac{q}{m_{\psi}}\Biggl\{\frac{g}{2}\begin{pmatrix}
0 \\
\mathbf{s}\times\mathbf{B} \\
\end{pmatrix}&+\left(1-\frac{g}{2}\right)\big(1-\widetilde{\chi}\big)[\mathbf{v}\times\mathbf{B}]\cdot\mathbf{s} \begin{pmatrix}
1 \\
\mathbf{v} \\
\end{pmatrix}\Biggr\} \notag \\
&=\frac{qB}{2m_{\psi}}\begin{pmatrix}
(1+2\chi)(g-2-4\chi)\widetilde{v}\sin(\Omega t) \\
g\cos[(\omega+\Omega)t] \\
g\sin[(\omega+\Omega)t] \\
0 \\
\end{pmatrix}+\mathcal{O}(\widetilde{v}^2)\,.
\end{align}
It can be checked that both sides of the BMT equation are equal for the following choice
\begin{equation}
\label{eq:rate-transverse-longitudinal}
\Omega=-(1-\widetilde{\chi})\frac{qB}{m_{\psi}}\left(\frac{g}{2}-1\right)=\frac{q}{m_{\psi}}\left(\frac{g}{2}-1\right)\big(1-\widetilde{\chi}\big)\widehat{\mathbf{v}}\cdot (\mathbf{B}\times\widehat{\mathbf{n}})\,.
\end{equation}%%
The structure of the latter result corresponds to Eq.~(9) in \cite{Bargmann:1959gz} for a vanishing electric field, but the global prefactor
is modified by Lorentz violation. This alters the rate at which the transverse spin component is transformed into the longitudinal one (and vice
versa). The modification is indirectly caused by the modified particle kinematics where the spin itself does not introduce any Lorentz-violating
effects at first order in Lorentz violation. Observing that the modified classical trajectory of \eqref{eq:circular-movement-magnetic-field-complete},
the BMT equation (\ref{eq:spin-eom-nonrelativistic}), and the modified rate of \eqref{eq:rate-transverse-longitudinal} are consistent with
each other at first order in Lorentz violation demonstrates that the used approach is reasonable.
Note that the kinematic term including Lorentz violation into the BMT equation comes with a minus sign. So the frequency $\Omega$
is reduced by a positive Lorentz-violating coefficient $\chi$. This behavior is in contrast to the increase of the frequency $\omega$.
Furthermore the modification is independent of the Land\'{e} factor as expected, since it originates purely from Lorentz-violating kinematics.

For isotropic Lorentz-violating frameworks the BMT equation of \eqref{eq:spin-eom} is expected to be modified at
second order in Lorentz violation. This modification is supposed to involve the timelike preferred spacetime direction $(\xi^{\mu})=(1,0,0,0)^T$
and it may lead to novel precession effects that are not governed by the \textit{Ansatz} of \eqref{eq:spin-ansatz}. The spin four-vector
itself would then involve $\xi^{\mu}$ as well and upon a particle Lorentz transformation the condition $s\cdot u=0$ would not be satisfied any
more. Therefore one possible signal of Lorentz violation could be that the spin four-vector does not stay perpendicular to the
four-velocity.

%---------------------------------------------------------------------------------------------------
\section{Discussion and outlook}
\label{sec:discussion}
%---------------------------------------------------------------------------------------------------

In this article the properties of a generic isotropic dispersion relation of the SME fermion sector were on the focus. The corresponding
classical, relativistic Lagrangian was determined and it was promoted to a Finsler structure. It was shown that the associated Finsler space
is Riemannian.

The classical particle was then assigned an electric charge and it was coupled to an external electromagnetic field. By doing so, the equations of
motion were determined and solved for particular cases.
The resulting particle trajectories were shown to be very similar to the standard ones with the difference that some quantities are scaled
due to the presence of the isotropic Lorentz-violating background field.

Subsequently the goal was to understand the behavior of particle spin. Since spin is a quantum theoretical concept, for the classical
particle it had to be introduced by hand. Its time evolution was derived by considering a modified version of the BMT equation. The result
is that for a nonvanishing magnetic field the rate is modified at which the transverse component is transferred into the longitudinal one and
vice versa. However at first order, Lorentz violation does not have any influence on spin precession in the magnetic field. A
modification is expected to occur for a Lorentz-violating theory exhibiting dispersion relations depending on spin projection.
However those dispersion relations were not considered here.

The paper shows that classical calculations within an isotropic, fermionic framework are feasible, which supports the consideration of isotropic
models first before delving into more complicated\footnote{Possible isotropic subspaces, e.g., of the b-structure may be treatable on the same level
of complexity, though.} frameworks based
on the b-structure, for example \cite{Kostelecky:2010hs,Kostelecky:2011qz}.
A reasonable conclusion is to associate the properties of the modified physics, e.g., scaled particle trajectories in electromagnetic fields
and a scaling of the transition rate between transverse and longitudinal spin components, with Euclidean Finsler structures.
A next step might be to promote the flat
intrinsic metric $r_{\mu\nu}$ to a curved metric $g_{\mu\nu}(x)$ and the constant coefficient $\Upsilon$ to a spacetime-dependent
function $\Upsilon(x)$. Studying the particle trajectories in such a spacetime may be a further step towards a better understanding of
Lorentz violation in the context of gravity.

Recently a paper has appeared on how to apply the concept of Finsler geometry to the photon sector~\cite{Itin:2014uia}. The
approach of the latter reference differs from what is carried out in the current article. A future purpose will be to apply our procedure to
the photon sector of the SME. It remains to be seen whether the results to be obtained will be consistent with the conclusions of \cite{Itin:2014uia}.

\section*{Acknowledgments}

It is a pleasure to thank D. Colladay and V.~A.~Kosteleck\'{y} for reading the manuscript and giving helpful suggestions. Furthermore the author is
indebted to the anonymous referee for helpful comments on the submitted version of the article. This work was performed with financial
support from the \textit{Deutsche Akademie der Naturforscher Leopoldina} within Grant No. LPDS 2012-17.

\newpage%%tmp
%---------------------------------------------------------------------------------------------------

%--------------------------------------------------------------------------------------------------

\end{document}